\documentclass[aps,prl,twocolumn,superscriptaddress]{revtex4}
\usepackage{graphicx}
\begin{document}

% Use the \preprint command to place your local institutional report
% number in the upper righthand corner of the title page in preprint mode.
% Multiple \preprint commands are allowed.
% Use the 'preprintnumbers' class option to override journal defaults
% to display numbers if necessary
%\preprint{}
%Title of paper
\title{Coexistence and competition of the short-range incommensurate antiferromagnetic order with superconductivity in BaFe$_{2-x}$Ni$_{x}$As$_{2}$}
\author{Huiqian Luo}
\affiliation{Beijing National Laboratory for Condensed Matter
Physics, Institute of Physics, Chinese Academy of Sciences, Beijing
100190, China}
\author{Rui Zhang}
\affiliation{Beijing National Laboratory for Condensed Matter
Physics, Institute of Physics, Chinese Academy of Sciences, Beijing
100190, China}
\author{Mark Laver}
\affiliation{Laboratory for Neutron Scattering, Paul Scherrer
Institute, CH-5232 Villigen, Switzerland}
\affiliation{ Department
of Physics, Technical University of Denmark, DK-2800 Kgs. Lyngby,
Denmark}
\author{Zahra Yamani}
\affiliation{Canadian Neutron Beam Centre, National Research
Council, Chalk River Laboratories, Chalk River, Ontario K0J 1J0,
Canada}
\author{Meng Wang}
\affiliation{Beijing National Laboratory for Condensed Matter
Physics, Institute of Physics, Chinese Academy of Sciences, Beijing
100190, China}
\author{Xingye Lu}
\affiliation{Beijing National Laboratory for Condensed Matter
Physics, Institute of Physics, Chinese Academy of Sciences, Beijing
100190, China}
 \affiliation{ Department of Physics and Astronomy,
The University of Tennessee, Knoxville, Tennessee 37996-1200, USA }
\author{Miaoyin Wang}
 \affiliation{ Department of Physics and Astronomy,
The University of Tennessee, Knoxville, Tennessee 37996-1200, USA }
\author{Yanchao Chen}
\affiliation{Beijing National Laboratory for Condensed Matter
Physics, Institute of Physics, Chinese Academy of Sciences, Beijing
100190, China}
\affiliation{Science and Technology on Nuclear Data
Laboratory, China Institute of Atomic Energy, Beijing 102413, China}
\author{Shiliang Li}
\affiliation{Beijing National Laboratory for Condensed Matter
Physics, Institute of Physics, Chinese Academy of Sciences, Beijing
100190, China}
\author{Sung Chang}
\affiliation{NIST Center for Neutron Research, National Institute of
Standards and Technology, Gaithersburg, Maryland 20899, USA}
\author{Jeffrey W. Lynn}
\affiliation{NIST Center for Neutron Research, National Institute of
Standards and Technology, Gaithersburg, Maryland 20899, USA}
\author{Pengcheng Dai}
\email{pdai@utk.edu}
\affiliation{ Department of Physics and Astronomy,
The University of Tennessee, Knoxville, Tennessee 37996-1200, USA }
\affiliation{Beijing National Laboratory for
Condensed Matter Physics, Institute of Physics, Chinese Academy of
Sciences, Beijing 100190, China}

% insert suggested PACS numbers in braces on next line


%\maketitle must follow title, authors, abstract, \pacs, and \keywords
\begin{abstract}
Superconductivity in the iron pnictides develops near antiferromagnetism,
and the antiferromagnetic (AF) phase appears to overlap with the
superconducting phase in some materials such as BaFe$_{2-x}T_{x}$As$_{2}$
(where $T=$ Co or Ni). Here we use neutron scattering to demonstrate that
genuine long-range AF order and superconductivity do not coexist in BaFe$%
_{2-x}$Ni$_{x}$As$_{2}$ near optimal superconductivity. In addition, we find
a first-order-like AF to superconductivity phase transition with no evidence
for a magnetic quantum critical point. Instead, the data reveal that
incommensurate short-range AF order coexists and competes with
superconductivity, where the AF spin correlation length is comparable to the
superconducting coherence length.
\end{abstract}

\maketitle

High-temperature superconductivity (high-$T_c$) in iron
pnictides arises at the border of antiferromagnetism
\cite{kamihara,cruz,jzhao}.  Since magnetic excitations may be
responsible for electron pairing and superconductivity
\cite{mazin,chubkov,fwang,jphu}, it is essential to understand the
doping and temperature dependence of the antiferromagnetic (AF) spin
correlations. For electron-doped iron pnictides such as
BaFe$_{2-x}T_{x}$As$_{2}$ (where $T=$ Co or Ni), the N${\rm \acute{e}}$el temperature
($T_N$) of the system decreases gradually with increasing
electron-doping level $x$ and the AF phase appears to overlap with
the superconducting phase \cite{nini,jhchu,clester}. This raises the
question concerning the role of quantum criticality \cite{abrahams}
and the coexisting AF order and superconductivity to the
superconducting pairing mechanism \cite{fernandes1,fernandes2}. Here
we use neutron scattering and transport measurements to show that
genuine long-range AF order does not coexist with superconductivity
in BaFe$_{2-x}$Ni$_{x}$As$_{2}$ near optimal doping. With increasing
$x$, the static AF order in BaFe$_{2-x}$Ni$_{x}$As$_{2}$ changes
abruptly from a commensurate wave vector for $x=0.085$
to an incommensurate wave vector with short-range order for $x=0.092, 0.096$.  While the ordered moment decreases smoothly from $x=0.085$ to 0.096,
the N${\rm \acute{e}}$el temperature ($T_N$) changes slowly from $\sim$47 K for $x=0.085$ to
$\sim$35 K for $x=0.096$ before vanishing at $x=0.1$.
 In addition, we find that the short-range incommensurate AF order
directly competes with superconductivity, and
there is no evidence
for a conventional magnetic quantum phase transition between the two phases.
Therefore, the presence of microscopic coexisting long-range AF and
superconducting phases and a magnetic quantum critical point (QCP) between
the AF and superconducting phase are not essential for
superconductivity in the BaFe$_{2-x}T_x$As$_2$ family of
materials.

In earlier neutron and X-ray scattering work on
BaFe$_{2-x}T_x$As$_2$,
the competition between coexisting superconductivity and antiferromagnetism
was inferred from the reduction of the
magnetic Bragg peak intensity below $T_c$
\cite{pratt,christianson09,mywang,kim,mywang11}.
If superconductivity and static long-range AF order coexist microscopically and compete for the same electrons, the superconducting pairing symmetry is most likely sign-reversed $s^{\pm}$-wave \cite{fernandes1,fernandes2}.
However,  muon spin rotation ($\mu$SR) experiments on underdoped
BaFe$_{1.89}$Co$_{0.11}$As$_2$ suggest an incommensurate spin density wave below $T^{mag}\approx 32$ K with a reduced ordered magnetic
moment below $T_c=21.7$ K \cite{marsik}. Neutron scattering reveals that
the commensurate AF order at the wave
vector $Q=(0.5,0.5,1)$ becomes transversely incommensurate at
$Q=(0.5-\delta,0.5+\delta,1)$ (inset in Fig. 1a) for
BaFe$_{2-x}$Co$_x$As$_2$ with $0.112<x<0.12$ \cite{pratt11}.

We carried out systematic neutron scattering
experiments on BaFe$_{2-x}$Ni$_x$As$_2$
using C-5, Rita-2, and BT-7 triple-axis
spectrometers at the Canadian Neutron Beam Center, Paul Scherrer
Institute, and NIST Center for Neutron Research (NCNR),
respectively. For C-5 and BT-7 thermal triple-axis spectrometers,
the final neutron energies were set to $E_f=14.56$ and $E_f=13.8$ meV, respectively,
with pyrolytic graphite (PG) as monochromator, analyzer, and filters.  For Rita-2
measurements,  the final energy was $E_f=4.6$ meV and a cooled Be filter was additionally used
as a filter. High quality single crystals were grown by FeAs
self-flux method as described previously \cite{Chen}. We define the
wave vector $Q$ at ($q_x$, $q_y$, $q_z$) as $(H,K,L) = (q_xa/2\pi,
q_yb/2\pi, q_zc/2\pi)$ reciprocal lattice units (rlu) using the
tetragonal unit cell, where $a \approx b \approx 3.96$ \AA, and $c =
12.77$ \AA.

\begin{figure}[t]
\includegraphics[scale=.35]{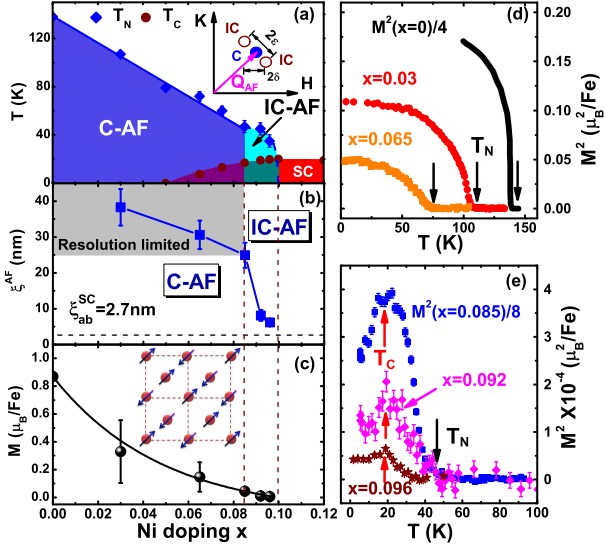}
\caption{(a) Electronic phase diagram of
BaFe$_{2-x}$Ni$_{x}$As$_{2}$ as a function of $x$. The
long-range commensurate AF (C-AF) order changes into short range
incommensurate AF (IC-AF) order for $x=0.085-0.096$.  The optimal
superconductivity occurs at $x=0.10$, where the static AF order is
suppressed \cite{schi}.  The inset shows positions of C-AF and IC-AF positions in
reciprocal space in tetragonal notation, where
$\delta=\epsilon/\sqrt{2}$. (b) The Ni-doping
dependence of the AF spin-spin correlation length. For $x$=0.096, we have $\xi^{AF} \approx 66$ {\AA} and the
superconducting coherence length $\xi^{SC}_{ab} \approx 27$ {\AA} \cite{yamatomo}.
(c) The doping dependence of the ordered magnetic moment $M$ \cite{note}. (d)
Temperature dependence of the
magnetic order parameter at $Q=(0.5, 0.5, 1)$ and (0.5, 0.5, 3) AF
Bragg positions for $x=0, 0.03, 0.065$, and (e) $x=0.085, 0.092, 0.096$.
 }
\end{figure}

Figure 1a shows the
electronic phase diagram of BaFe$_{2-x}$Ni$_x$As$_2$ as a function
of Ni-doping $x$ as determined from our neutron scattering
experiments, where the commensurate to incommensurate AF phase
transition occurs between $x=0.085$ and 0.092.
Figures 1b, 1c, 1d and 1e show the Ni-doping dependence of the spin correlation length, moment, and magnetic order parameters, respectively.
While the
N${\rm \acute{e}}$el temperatures decrease gradually with increasing
$x$ for $0\le x\le 0.065$ as shown in Fig. 1d \cite{pratt,christianson09,mywang,kim,mywang11},
they decrease rather slowly for $x=0.085, 0.092, 0.096$ before
vanishing abruptly at $x=0.1$ (Figs. 1a and 1e) \cite{schi}.
For comparison,
the ordered moment decreases smoothly  to zero with increasing $x$
at 0.1 (Fig. 1c), consistent with the presence of a magnetic QCP \cite{abrahams}.
However, the static spin correlation length, which is instrumental resolution limited for $x<0.085$,
decreases abruptly for samples with incommensurate AF order (Fig. 1b) \cite{note}.
This is contrary to the magnetic QCP in CeFeAs$_{1-x}$P$_x$O
where the commensurate AF order is resolution limited at all $x$ as the ordered moment
vanishes with $x\rightarrow 0.4$ \cite{delacruz}.

\begin{figure}[t]
\includegraphics[scale=.35]{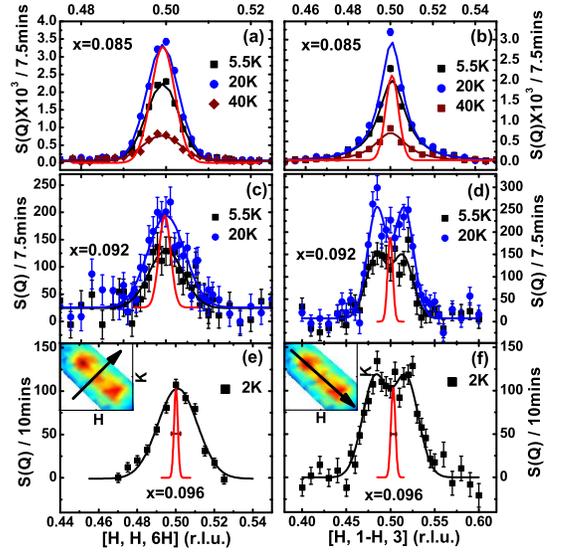}
\caption{Temperature and wave vector dependence of the C-AF and IC-AF scattering for BaFe$_{2-x}$Ni$_x$As$_2$ with $x=0.085, 0.092$ and 0.096. Samples are aligned in the $[H,H,6H]$ and $[H,1-H,3]$ scattering plane.
The solid red lines are the instrumental resolutions obtained using $\lambda/2$ scattering from the
 $(1, 1, 6)$ nuclear Bragg peak above $T_N$ without filter. Data in (a-d) are collected on C-5.
 (a,b) Longitudinal and transverse scans at different temperatures through the $(0.5,0.5,3)$ AF Bragg peak for $x=0.085$.  The scattering is commenusrate in both directions but not instrumental resolution limited.
The solid lines in transverse scans are Lorentzian fits to the data.
(c,d) Identical scans using the same experimental setup for $x=0.092$, which show clear incommensurate scattering along the transverse direction.
(e,f) Longitudinal and transverse scans for
 $x=0.096$ at $Q=(0.5, 0.5, 3)$ collected on Rita-2.
 The solid horizontal bars are the calculated instrumental resolution,
 determined by the supermirror guide before the monochromator, the $80^\prime$ collimation,
 the radial collimator of the Be filter (about $150^\prime$), the neutron absorbing
 guide after the analyzer (effective collimation of $40^\prime$)
 and a sample mosaic spread of $\sim$$15^\prime$.  Inserts show the color images of incommensurate peaks centered around $Q=(0.5, 0.5, 3)$ and the scan directions at 2 K.
 }
 \end{figure}

To demonstrate the doping evolution of the AF order through the commensurate to incommensurate AF phase transition,
we summarize in Fig. 2 longitudinal and transverse
scans along the $[H,H,6H]$ and $[H,1-H,3]$ directions for $x=0.085, 0.092$, and 0.096 at different temperatures, where the solid red lines indicate
the instrumental resolution.
For $x=0.085$, the scattering is commensurate along both the longitudinal and transverse directions, but not instrumental resolution limited (Figs. 2a and 2b).
Furthermore, the lineshape of transverse scan is not a Gaussian but can be fit with a Lorentzian.
Figures 2c and 2d show identical scans using the same
instrument for the $x=0.092$ sample.  Here, we find broad commensurate scattering in the longitudinal direction and clear incommensurate peaks in the transverse direction.
 Figures 2e and 2f plot longitudinal and transverse
scans along the aforementioned directions for the $x=0.096$ sample.
Converting these widths into real space \cite{note}, we find that the static spin correlation
length along the longitudinal direction is only $62\pm 5$ \AA\,
while it is $81\pm 15$ \AA\ and $249\pm 35$ \AA\ for the $x=0.092$
and 0.085 samples, respectively.

\begin{figure}[t]
\includegraphics[scale=.35]{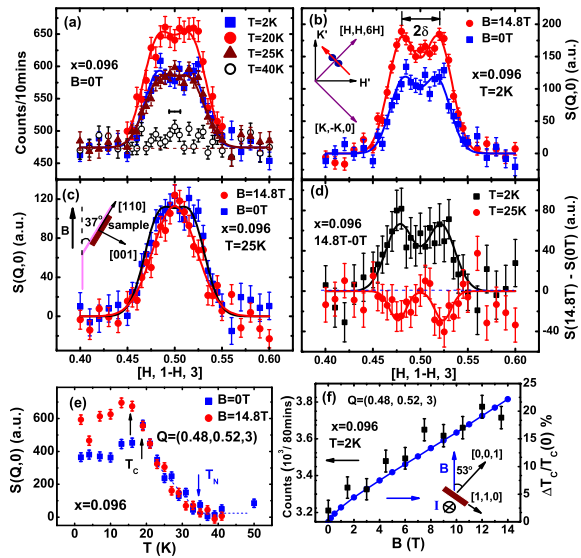}
\caption{
 Effect of a 14.8 T magnetic field on the short-range incommensurate AF order in the
 $x=0.096$ sample. For this experiment on Rita-2, a single crystal of mass about 0.5 gram and mosaic of
 0.45$^\circ$ was aligned in a 15-T magnet in the $[H,H,6H]$ and $[H,1-H,3]$
 scattering plane as shown in the inserts of (b) and (c).
 Neutrons of $E_i=E_f= 4.6$ meV were selected,
 with the nine blades of the analyzer set to probe different points in reciprocal space.
Two filters were employed to remove higher order neutrons: a pyrolytic
graphite placed just before the sample, along with a $80^\prime$ collimator,
and cooled Be placed just after the sample. (a) Transverse
scans at zero field and at $T = 2$ K (below
 $T_c$), 20 K (around $T_c$) , 25 K ($T_c < T < T_N$) and 40 K (above $T_N$).  The incommensurability $\pm\delta$ remains at 0.018 for all measured
 temperatures. (b) Comparison between zero and 14.8 T fields at $T=2$ K. The 14.8 T magnetic field clearly
 enhances the IC-AF order. (c) Identical scans at $T=25$ K, where a 14.8 T field
 suppresses the incommensurate AF order. (d) The effect of a 14.8 T field on the
incommensurate AF order.  While the
 field enhances the IC-AF order at 2 K, it may suppress the IC-AF order in the normal state.
 (e) Temperature dependence of the magnetic order parameters at the incommensurate position
$Q=(0.48, 0.52, 3)$ for 0 and 14.8 T fields. The superconducting transition temperature $T_c$
is seen to shift from $\sim$19 K to $\sim$15 K.
 (f) Magnetic field dependence of the AF Bragg peak intensity at
 $Q=(0.48,0.52,3)$.  The data show a linear field dependence,
consistent with a field-induced suppression of superconducting transition
temperature $\Delta T_c=(T_c(0)-T_c(B))/T_c(0)$ as determined from resistivity measurements on the
same sample (solid blue circles and lines).
 }
\end{figure}

Figure 3a-3c shows the detailed temperature dependence of the
transverse scans at zero and a field of 14.8 T. At zero field,
transverse scans are featureless at $T=40$~K ($>T_N$) but show broad
peaks indicative of incommensurate AF short-range order below
$T_N$$\approx$ 35 K. At $T=20$ K just above $T_c$, the peak
intensity continues to increase, but decreases upon further cooling
to 2 K (Fig. 3a). These results are consistent with earlier work
on BaFe$_{2-x}$Co$_x$As$_2$ \cite{pratt11}. Upon applying a 14.8 T
field aligned at $\sim$37$^\circ$ out of the FeAs-plane (Fig. 3b), we
see that the broad peak at zero field and 2 K increases in intensity
and becomes two clear incommensurate peaks centered at
$Q=(0.5-\delta,0.5+\delta,3)$ with $\delta=0.018\pm 0.002$ rlu.
For a temperature just above $T_c$ at 25 K, the broad peaks appear to
merge into a single commensurate peak centered at $Q=(0.5,0.5,3)$
(Fig. 3c). To determine the net effect of a 14.8 T field, we show in
Fig 3d the field-on minus field-off difference plots.  At $T=2$ K,
the effect of a field is to induce clear incommensurate peaks,
different from the field effect on superconducting
BaFe$_{2-x}$Ni$_x$As$_2$ with lower $x$ \cite{mywang11}. At a
temperature ($T=25$ K) just above $T_c$, the effect of a field
appears to be opposite and suppresses the incommensurate AF order.
Figure 3e shows the temperature dependence of the scattering at the
incommensurate position at zero and 14.8 T. At zero field, the data
reveal a clear suppression of the magnetic intensity at $T_c$.  A
14.8 T field reduces $T_c$ from 19 K to 15 K and enhances the
incommensurate AF order. The intensity of the incommensurate AF
scattering increases linearly with increasing field, consistent with
the field-induced reduction in the superconducting transition
temperature as determined from resistivity measurements (Fig. 3f).
However, the linewidths of the incommensurate peaks remain unchanged at 2 K (Fig. 3b).  Therefore,
superconductivity competes with the short-range incommensurate
AF order instead of the long-range AF order.

\begin{figure}[t]
\includegraphics[scale=.25]{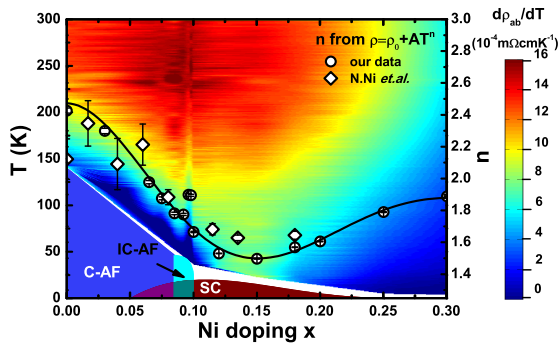}
\caption{ Temperature and doping dependence of in-plane resistivity of BaFe$_{2-x}$Ni$_x$As$_2$
in the normal state ($T>T_N,T>T_c$) derived from ref. \cite{Chen}
and ref. \cite{NNi}, where $\rho_{ab}(T=300\ {\rm K})$ is assumed to
decrease linearly with increasing $x$. The gradient color is the first order
differential of resistivity in the normal state, $d\rho_{ab}/dT$,
suggesting the linear term emerges from the overdoped regime. Open
symbols are the exponent $n$ deduced from fitting $\rho=\rho_0+AT^n$
in the normal state, which minimizes to $n\approx 1.5$ near
$x=0.15$.  An anomalous $n$ is found for the IC-AF sample.
 }
\end{figure}

In transport and nuclear magnetic resonance (NMR) experiments on
isoelectronic BaFe$_2$(As$_{1-x}$P$_x$)$_2$
 \cite{sjiang,skasahara,ynakai,ynakai1},
a magnetic QCP has been identified at $x=0.33$,
which is believed to play an important role in the superconductivity
of these materials \cite{abrahams}.  For
BaFe$_{2-x}$Co$_{x}$As$_{2}$, recent systematic ultrasonic
measurements \cite{yoshizawa} suggest the presence of a structural
QCP near optimal superconductivity, where the
structural distortion associated with the static AF order vanishes.
These results are consistent with NMR measurements, where the
strength of the paramagnetic spin fluctuations diverges for Co
concentration near optimal superconductivity \cite{ning}. If
Ni-doping in BaFe$_2$As$_2$ is equivalent to twice the Co-doping
\cite{budko}, one should also expect a structural and magnetic
QCP near optimal superconductivity for
BaFe$_{2-x}$Ni$_{x}$As$_{2}$.
Since the incommensurate AF spin
correlations for the $x=0.092, 0.096$ samples clearly do not increase with
decreasing temperature,
it is difficult to reconcile this result
with a magnetic QCP, where one expects a
diverging spin-spin correlation length as $T\rightarrow 0$ K.  Furthermore,
the N${\rm \acute{e}}$el temperature of BaFe$_{2-x}$Ni$_{x}$As$_{2}$
 suddenly vanishes at $x=0.1$ from $T_N\sim$35 K for $x=0.096$.  Therefore, instead of a magnetic QCP,
 the incommensurate AF order to superconductivity phase transition in
 BaFe$_{2-x}$Ni$_{x}$As$_{2}$ appears to be a first order,
 much like that of the LaFeAsO$_{1-x}$F$_x$ family of materials \cite{leutkens}.

If there is a magnetic QCP in the phase diagram
of BaFe$_{2-x}$Ni$_{x}$As$_{2}$ near $x=0.10$ where the static
long-range AF order vanishes \cite{schi}, the temperature dependence
of the resistivity $\rho=\rho_0+AT^n$ should have an exponent
$n\approx 1$ near $x=0.1$ within a single band model similar to that of
BaFe$_2$(As$_{1-x}$P$_x$)$_2$ at $x=0.33$ \cite{ynakai1}. Figure 4
shows the electron-doping dependence of the resistivity exponent $n$
obtained by fitting the temperature dependence of the resistivity of
BaFe$_{2-x}$Ni$_{x}$As$_{2}$ \cite{Chen,NNi}. The resistivity
exponents show a broad minimum with $n\approx 1.5$ near $x=0.15$.
Similar analysis on the in-plane resistivity data of BaFe$_{2-x}$Co$_x$As$_2$ in the normal state also yielded
minimum $n$ in the overdoped region, clearly different from that for
BaFe$_2$(As$_{1-x}$P$_x$)$_2$ \cite{ynakai1}.
Therefore, our data suggest no magnetic QCP near the boundary of
AF and superconducting phases in BaFe$_{2-x}T_x$As$_2$.  This is consistent with the more
accurate two band analysis of the normal state resistivity for BaFe$_{2-x}T_x$As$_2$
\cite{dwu,nbarisic}, where a Fermi liquid like coefficient $n=2$ was found for optimally doped
BaFe$_{2-x}T_x$As$_2$ again suggesting no QCP near optimal superconductivity.

The observation of competing static short-range incommensurate AF
order with superconductivity and the first-order-like
AF to superconductivity phase transition
raises the question concerning
how AF order microscopically coexists with superconductivity in
Fe-based superconductors \cite{mywang11}. In a recent $^{57}$Fe
M${\rm \ddot{o}}$ssbauer spectroscopy study of
BaFe$_{2-x}$Ni$_{x}$As$_{2}$, a small reduction in magnetic
hyperfine field below $T_c$ was found for the $x=0.085$ sample,
suggesting coexisting AF order and superconductivity on a length
scale of $\sim$27 \AA\  \cite{munevar}. However, our data show that AF order at this doping level is commensurate with a correlation length of $\sim$250 \AA\ (Figs. 1b and 2a).
For hole-doped
Ba$_{1-x}$K$_x$Fe$_2$As$_2$, $\mu$SR \cite{wiesenmayer}
and neutron powder diffraction \cite{hchen,avci} measurements have
suggested microscopic coexisting AF and superconducting phases in
the underdoped regime.
However, these measurements did not probe the region of the phase diagram close to optimal superconductivity, and were unable to provide a length scale for the AF order that coexists with the superconductivity.
 From Figs. 1-3, we see that the static
incommensurate AF order competing with superconductivity has a
spin-spin correlation length of $\sim$60 \AA.  This means that the
incommensurate AF order has a similar length scale to the
superconducting coherence length ($\sim$27 \AA) \cite{yamatomo},
and that, near optimal doping, there is no long-range AF order coexisting with superconductivity.
Instead, our data can be understood in two scenarios:
first, the two orders coexist microscopically and homogeneously, and compete for the same itinerant
electrons \cite{fernandes1,fernandes2}, such that superconductivity occurs at the expense of the static AF order.
When a magnetic field is applied, the superconducting gap $\Delta(B)$
and $T_c$ decrease with increasing field via $\Delta(B)/\Delta(0)=
T_c(B)/T_c(0)=\sqrt{1-B/B_{c2}}$  \cite{jzhao10}. In the low-field limit, we have
 $B/B_{c2}\propto \Delta T_c/T_c(0)$.  Therefore, the field-induced AF order should be proportional to
 the field-induced reduction in $T_c$, consistent with the
data in Fig. 3f.
Alternatively, the competition is mesoscopic:
phase-separation occurs with superconducting and non-superconducting, AF-ordered nano-regions of length scale $\sim$60 \AA. In this picture, the superconducting electrons do not directly contribute to the static AF order, and superconductivity only affects the AF order through a proximity effect.  Here, one can imagine that the field-induced non-superconducting vortices have incommensurate AF order,
much like field-induced AF vortices in some
copper oxide superconductors \cite{lake}.  This is also consistent with the first-order-like AF to superconductivity transition with increasing $x$.
Since our neutron diffraction measurements of the bulk of the sample cannot resolve superconducting from non-superconducting parts of the sample, we find both scenarios are consistent with our observation of short-range AF order with superconductivity near optimal doping.

We thank Jiangping Hu, Qimiao Si, Daoxin Yao, Hai-Hu Wen, and Xingjiang Zhou
for helpful discussions and transport measurements. The work at IOP, CAS, is
supported by MOST (973 project: 2011CBA00110,
2010CB833102, and 2012CB821400) and NSFC (No.11004233). The work at UTK is supported by the
U.S. NSF-DMR-1063866 and NSF-OISE-0968226. Part of the work is based
on experiments performed at the Swiss Spallation Neutron Source,
Paul Scherrer Institute, Villigen, Switzerland.  M. L.
acknowledges support from DanScatt.

% Create the reference section using BibTeX:
%\bibliography{NoEndingPoint}

\end{document}

\appendix

\title{Supplemental Materials}

\makeatletter \renewcommand{\thefigure}{S\@arabic\c@figure} \renewcommand{\thetable}{S\@arabic\c@table} \makeatother
\setcounter{figure}{0}

\maketitle

\section{Coexistence and competition of the short-range incommensurate antiferromagnetic order with superconductivity in BaFe$_{2-x}$Ni$_{x}$As$_{2}$}

\begin{figure}
\begin{center}
\includegraphics[width=0.5\linewidth]{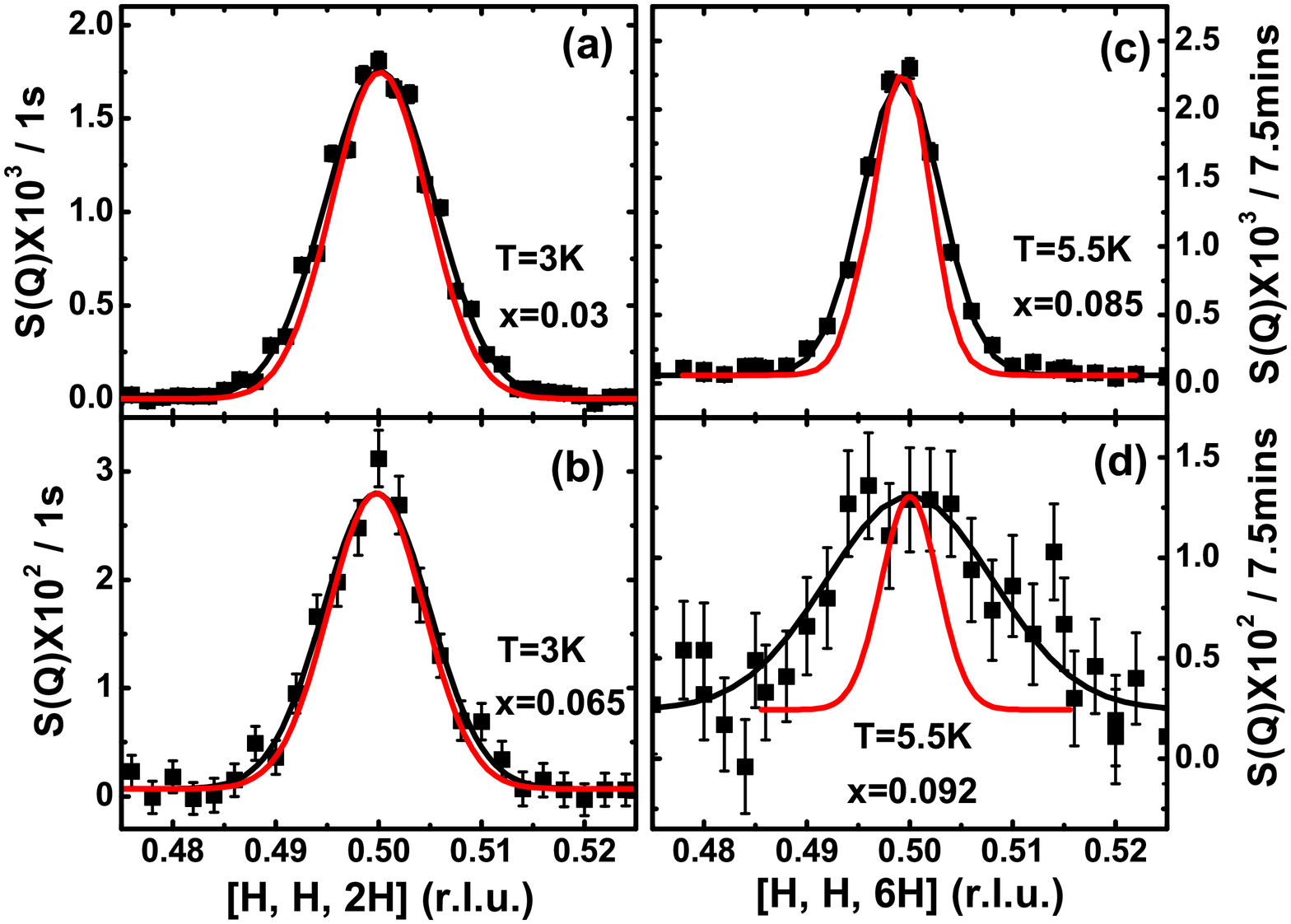}
\end{center}\caption{ (a,b)
Wave vector scans along the $[H,H,2H]$ direction for $x=0.03,0.065$
at 5 K.  The red solid lines are instrumental resolution obtained by
$\lambda/2$ scattering with filters above $T_N$.  The peak width is
resolution limited (c,d) Radial scans along the $[H,H,6H]$ direction
for $x=0.085,0.092$ at 5.5 K. The peak width is broader than the
instrumtal resolution.  The solid lines are Gaussian fits to the
data on linear backgrounds. In all cases, data were obtained by
subtracting low-temperature data from  background scattering above
$T_N$. } \label{Fig:figS1}
\end{figure}

\begin{table}[h]
\begin{center}
     \begin{tabular}{ |l  |l |l | l | l | l | l | l | }
     \hline
           BaFe$_{2-x}$Ni$_x$As$_2$ &Magnetic  &Normalized by & $T_N$     & Peak width       & Resolution       & Correlation         & Ordered  \\
 Composition &Peak  &Nuclear Peaks & (K)    & FWHM (rlu)       &FWHM (rlu)       &Length (\AA)         &Moments ($\mu_B$)  \\

 \hline $x=0.03$ &(0.5, 0.5, 1) &(1, 1, 0), (1, 1, 2) & $107(2)$ & $0.0124(2)$ &
$0.0109(2)$ &
$383(51)$ & $0.33(23)$ \\
\hline
$x=0.065$ &(0.5, 0.5, 1) &(1, 1, 0), (1, 1, 2) & $72(3)$ & $0.0122(5)$ & $0.0097(1)$ & $306(40)$ & $0.15(10)$ \\
\hline
$x=0.085$ &(0.5, 0.5, 3) &(1, 1, 0), (1, 1, 6) & $47(5)$ & $0.0090(3) $ & $0.0067(1)$ & $249(35)$ & $0.045(27)$ \\
\hline
$x=0.092$ &(0.5, 0.5, 3) &(1, 1, 0), (1, 1, 6) & $45(5)$ & $0.0195(21)$ & $0.0065(1)$ & $81(15)$ & $0.011(7)$ \\
\hline
$x=0.096$ &(0.5, 0.5, 3) &(1, 1, 0) & $35(5)$ & $0.0242(12)$ & $0.0023(1)$ & $62(5)$ & $0.0065(16)$ \\
\hline
     \end{tabular}
     \caption{\label{tab:5/tc}
The Ni-doping evolution of the N$\rm \acute{e}$el temperatures,
full-width-half-maximum (FHWM) of the Gaussian widths in rlu,
instrumental resolution obtained from $\lambda/2$ scattering of
Bragg peaks above $T_N$ without filter, estimated spin correlation
length, and ordered moments.
     }

 \end{center}
\end{table}

Figure S1 shows low-temperature longitudinal scans used to estimate
the spin correlation length as a function of Ni-doping $x$.  The
doping dependence of the width and errors are shown in Table S1. The
full-width-half-maximum (FHWM) is obtained by Gaussian fits to the
peaks on linear backgrounds, where $I_M=bkg +
(I_0/(w\sqrt{\pi/2}))e^{-2((H-H_c)/w)^2}$, here we define
FWHM$=W=\sqrt{2\ln2}w$. By Fourier transform of the Gaussian peaks
along $Q=[H, H, L]$, we can deduce the spin-spin correlation length
$\xi=8\ln2/(\sqrt{W^2-R^2}\mid Q_{HHL}\mid)$, where $R$ is the
resolution and $Q_{HHL}=2\pi\sqrt{(H/a)^2+(H/b)^2+(L/c)^2}$ with
lattice parameters $a \approx b \approx 3.96$ \AA, and $c = 12.77$
\AA. In our case, we have $\mid Q_{116}\mid=3.71/$\AA\ and $\mid
Q_{112}\mid=2.45/$\AA.

The magnetic moment is estimated by comparing the nuclear peak
intensity $I_N=AN_N(2\pi)^2/V_N \times(\mid
F_N(G)\mid)^2/\sin(2\theta_N)$ and the magnetic peak intensity
$I_M=AN_M(2\pi)^2/V_M \times(\mid F_M(G)\mid)^2/(2\sin(2\theta_M))$
with twinning, where the number of atoms in magnetic unit cell is
$N_M=N_N/2=4$, the volumes of magnetic unit cell $V_M=2V_N$,
$F_N(G)$ and $F_M(G)$ are the structure factor and magnetic form
factor at wave vector $\mathbf{G}$ with scattering angle $2\theta_N$
and $2\theta_M$, respectively. Here we have $F_M(G)=p\mid
S_{\perp}\mid\sum(-1)^ie^{iGd}$, where
$p=0.2659\times10^{-12}cm\times g\times f_M(G)$ with $Fe^{2+}$ form
factor
$f_M(G)=Ae^{aG^2/16\pi^2}+Be^{bG^2/16\pi^2}+Ce^{cG^2/16\pi^2}+D$,
$S_{\perp}$ is the magnetic moment along wave vector $\mathbf{G}$,
$d$ is the spacing of wave vector $\mathbf{G}$, and $g$ factor is
assumed to be 2. Thus we have $\mid S_{\perp}\mid
=0.0665\times\sqrt{I_M\sin{2\theta_M}/I_N\sin{2\theta_N}}\times \mid
F_N(G)\mid/\mid f_M(G)\mid$, and finally we obtain magnetic moment
per $Fe^{2+}$: $S=\mid S_{\perp}\mid/\sqrt{1-\cos^2\eta}$, where
$\eta$ is the angle between measured wave vector $\mathbf{G}$ and
ordered moment $\mathbf{S}$ which is along $Q=[H, H, 0]$.